\documentstyle[sprocl,epsf,epsfig]{article}

\def\Journal#1#2#3#4{{#1} {\bf #2}, #3 (#4)}

\def\PRL{\em Phys. Rev. Lett.}
\def\PRD{{\em Phys. Rev.} D}

\def\be{\begin{equation}}
\def\ee{\end{equation}}
\def\bea{\begin{eqnarray}}
\def\eea{\end{eqnarray}}
\newcommand {\ebar}{\hbox{E\kern-0.5em\lower-0.1ex\hbox{/}}}

\begin{document}

\title{NEW OBSERVATIONS OF TOP AT CDF}

\author{ SANDRA LEONE (for the CDF Collaboration)}
\address{I.N.F.N. Sez. di Pisa\\
Via Vecchia Livornese, 1291\\
I-56010 S.Piero a Grado (Pisa), Italy}


\maketitle\abstracts{We report preliminary results from the analyses
looking for top candidates in the channel with 6 jets in the final 
state (all hadronic) and in the dilepton channel with one tau lepton 
($\tau$ dilepton), using a data sample 
of about 110 $pb^{-1}$ collected by the CDF experiment. }

\section{Introduction}
After the first direct evidence for the top quark presented by CDF in 
1994~\cite{firstcdf}, 
the top quark discovery was announced by the CDF~\cite{cdftop} 
and D0~\cite{d0top} 
Collaborations in the early 1995. The channels with at least one lepton 
(electron $e$  
or muon $\mu$) in the final state have been used for those studies. 

After the top discovery, CDF focused to obtain a better understanding of the 
top quark 
properties and searched for a top signal in the other decay channels. 
This paper describes some preliminary studies that isolated a top signal 
in the all hadronic mode and in the dilepton channel with a $\tau$ lepton 
in the final state. A data sample of $\approx$ 110 $pb^{-1}$ has been used,
corresponding to the full statistics collected between 1992 and 1995.

In the all hadronic channel both $W$'s from top decay to hadrons: 
$$t\bar{t}\rightarrow W^+W^-b\bar{b} \rightarrow 
(q\bar{q}\prime)b(q\bar{q}\prime)\bar{b}.$$ 
The 
branching ratio is the largest of all top decays ($\approx$ 44\%). 
The final state has a 6 jets topology, with two jets coming 
from $b$ quarks. In principle the two top quarks could be
fully reconstructed because there are no neutrinos in the event. However, 
there is a huge QCD multijet background, 
orders of magnitude bigger than the signal, 
which includes real heavy flavor production through various processes.

Concerning the dilepton channel where one $W$ decays to a $\tau$ 
lepton and the other one to an $e$ or $\mu$, it 
represents 5\% of the total $t\bar{t}$ decay, exactly like the standard 
dilepton 
channels originally used for top search. We look for $\tau$'s in their hadronic 
decay: 
$$t\bar{t} \rightarrow W^+W^-b\bar{b} \rightarrow e(\mu)\nu_{e(\mu)}\tau
\nu_{\tau}b\bar{b},~\tau\rightarrow hadrons + \nu_{\tau}.$$
%Also this channel is challenging,
Compared to the other dilepton modes, this channel has a considerable 
background from jets misidentified as $\tau$'s.

\section{All hadronic channel}

The goal of the analysis in the all hadronic channel is to drastically reduce 
the background while 
keeping an efficiency as high as possible for $t\bar{t}$ detection
~\cite{azzitesi}. The strategy  
consists in first making a kinematic selection and then requiring the  
presence of $b$ quarks in the events. The starting sample is made of 
approximately
230,000 events, collected by a trigger requiring at least 4 jets 
reconstructed in a cone of radius $R$ = 0.4 in the $\eta - \phi$ space, with 
transverse energy $E_T$ $\ge$ 15 $GeV$ and pseudorapidity $|\eta|$$\le$ 2.
The signal over background ratio ($S/B$) at the trigger level is about 1/1000. 

\begin{figure}[th]
\centerline{
%\hskip 1.0cm
\epsfysize 6.35cm
\epsffile[24 149 549 689]{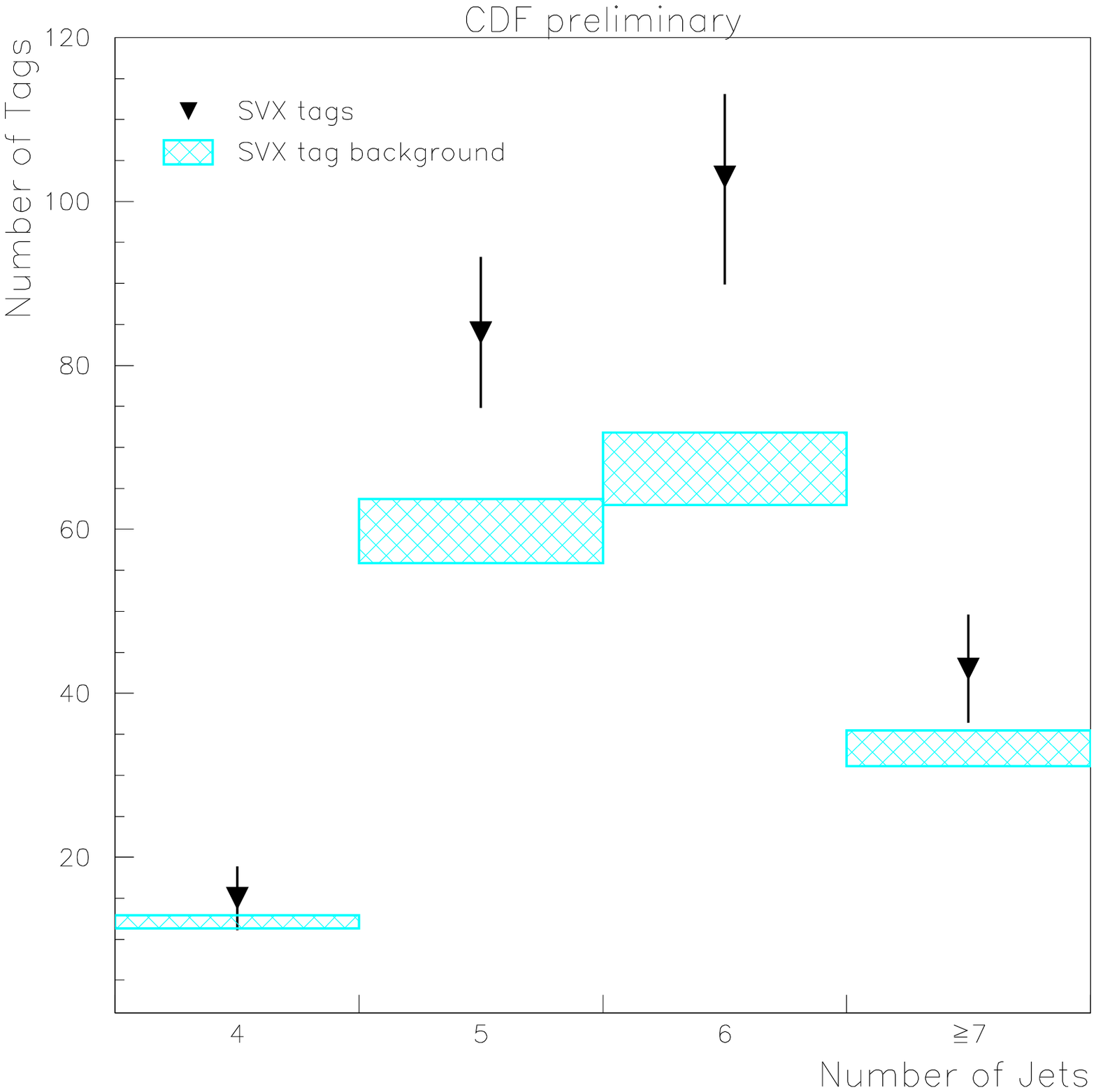}
\hskip -0.3cm
\epsfysize 6.35cm
\epsffile[24 149 549 689]{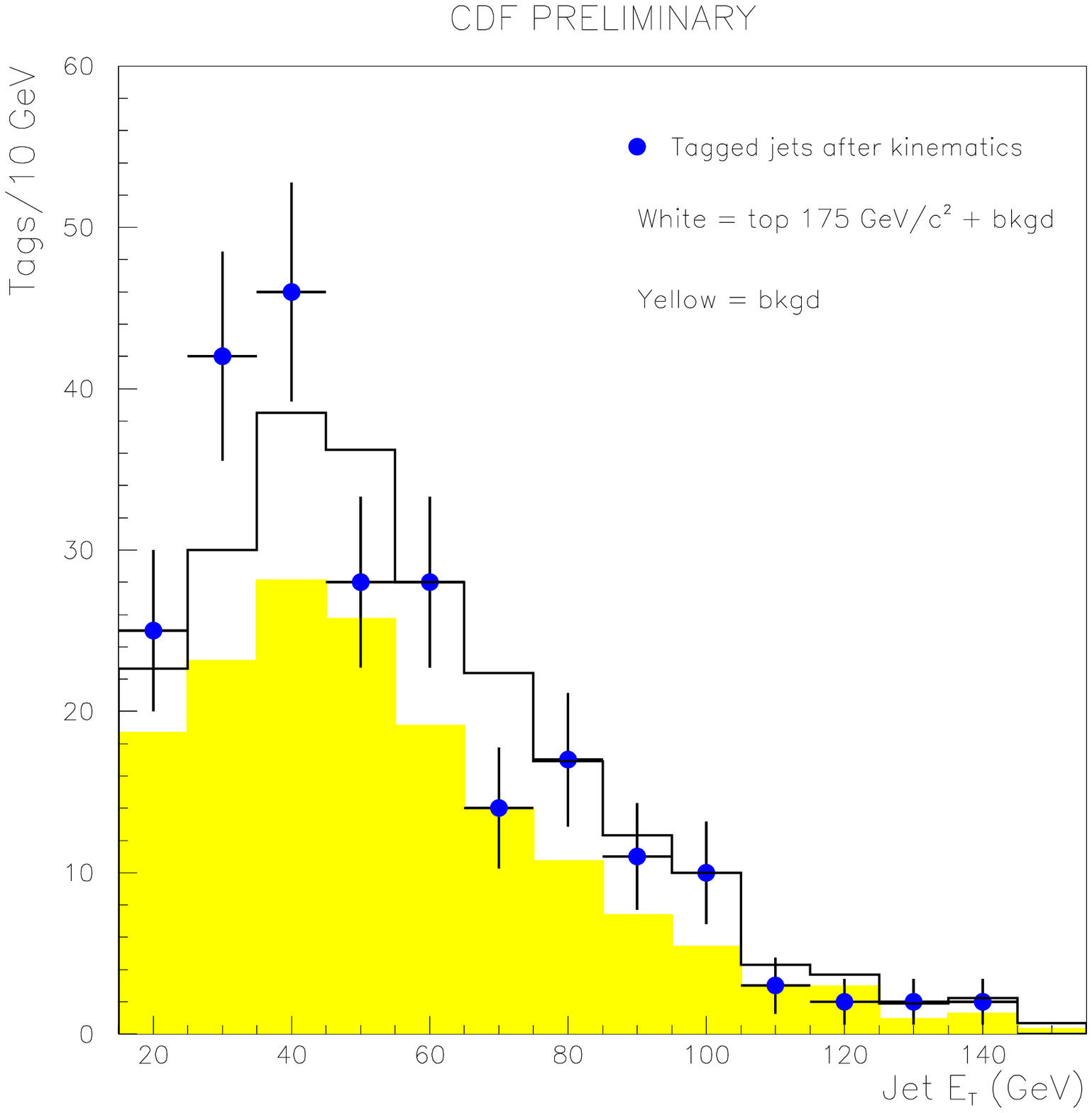}}
\vspace{-0.2cm}
\caption{Left (L): Number of $SVX$ tags as a function of jet multiplicity, 
compared with the expectations from background. 
Right (R): $b$--tagged jet $E_T$ distributions.}
\label{all_had}
\end{figure}

\vspace{-0.5cm}

\subsection{Kinematic selection and $b$ tag}

Top signal events are expected to have a higher jet multiplicity, to
be more central and to have a larger aplanarity than QCD background 
events. 
The selection starts by requiring a high jet multiplicity ($N(jets)$ $\ge$ 5).
We then cut on some global calorimetric variables like the total transverse 
energy of jets ($\sum E_T \ge$ 300 $GeV$), the fraction of transverse 
energy ($\sum$ $E_T$/$\sqrt{\hat{s}}$ $\ge$ 0.75) and the 
aplanarity 
($Apl$ $\ge$ $-0.0025~\times~\sum_{3}^{N}~E_T~+~0.54$, where the sum 
does not include the contribution from the 2 leading jets). The cuts have 
been chosen at the values which maximize the $S/B$ significance. After this 
kinematic selection, $S/B$ $\approx$ 1/30. The resulting data sample is 
still dominated by multijet production from QCD processes.

The requirement of at least one 
secondary vertex ($SVX$) $b$ tag~\cite{firstcdf} helps to further 
increase $S/B$ and extract 
a signal. Besides from $t\bar{t}$ events, $b$--tags can come from real 
heavy flavor production and from 
tracking mismeasurements. The tagging rate, defined as the number of tagged 
jets divided by the number of $taggable$ jets~\cite{firstcdf} has been 
parametrized as a function of $E_T(jet)$, $\eta(jet)$,
number of $SVX$ tracks associated with a jet and $Apl(jet)$. 
It has been calculated using a sample of events collected with a multijet 
trigger. This 
parametrization is found to describe well the jet multiplicity 
distribution of observed tags in multijet events. 

The sample selected with these characteristics consists of 192 events, 
containing 230 $b$--tagged jets. The number of tagged jets expected from 
background is 160.5 $\pm$ 10.4 tags. Fig.~\ref{all_had}(L) shows the number 
of tags as a function of jet multiplicity. The rectangles represent the 
background estimate from tag rate parametrization with systematic and 
statistical uncertainties. There is an excess of tagged jets in the jet 
multiplicity bins $N=5,6$ and $\ge 7$. The significance of the excess is 
estimated from the probability that the background fluctuates up to the 
number of found tags or greater. We found a ${\cal P}= 1.5\times 10^{-4}$,
corresponding to a 3.6$\sigma$ for a gaussian distribution (the effect of 
multiple tags is included in the calculation). Fig.~\ref{all_had}(R) shows the 
comparison between the $E_T$ spectra of the tagged jets in the data (black 
dots) and of the 
estimated background (shaded histogram). The white histogram is the sum of 
the background and $t\bar{t}$ contribution. 

\begin{figure}[tbh]\centerline{
%\hskip 1.0cm
\epsfysize 6.3cm
\epsffile[24 149 549 689]{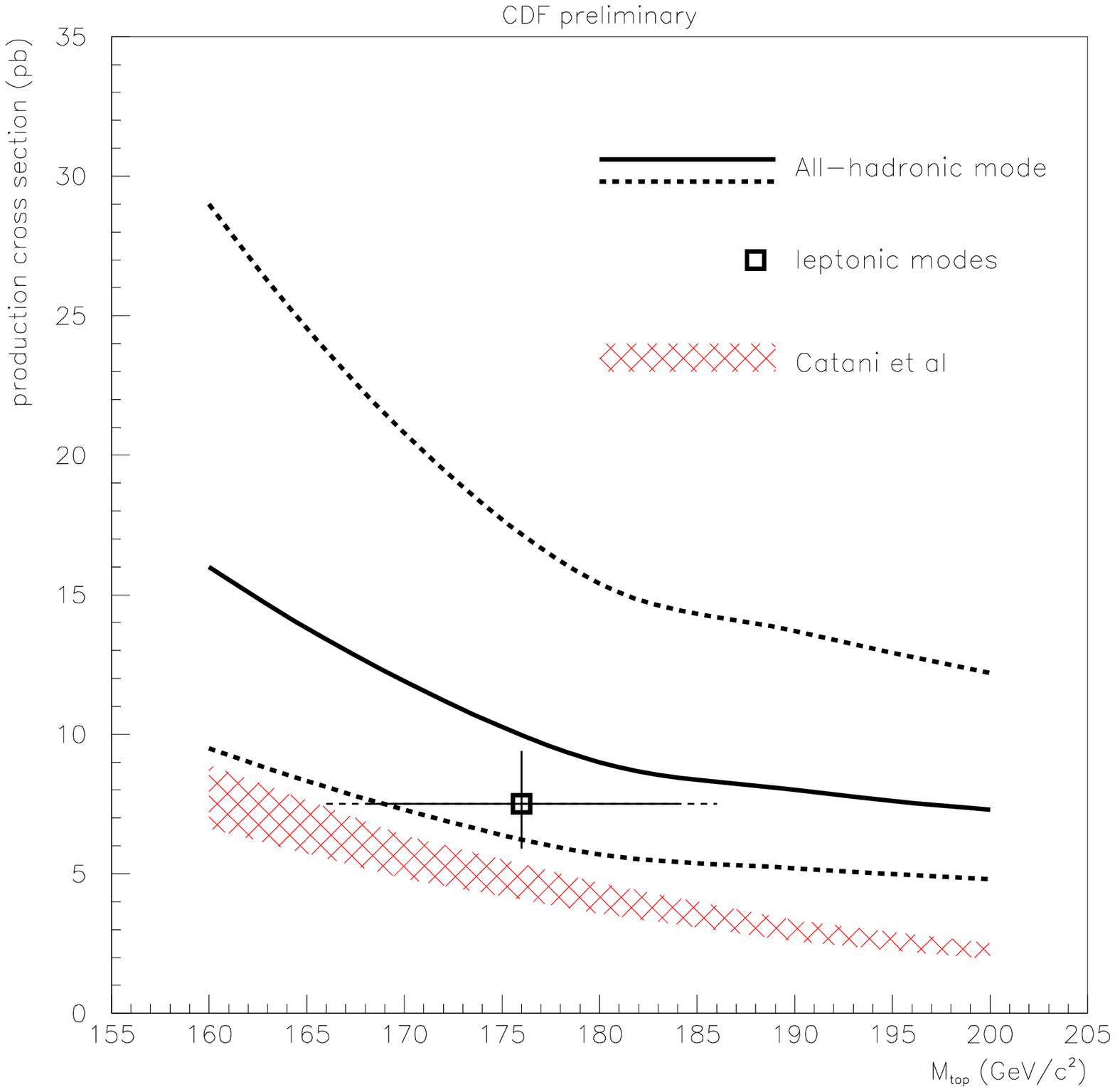}
\hskip -0.2cm
\epsfysize 6.3cm
\epsffile[24 149 549 689]{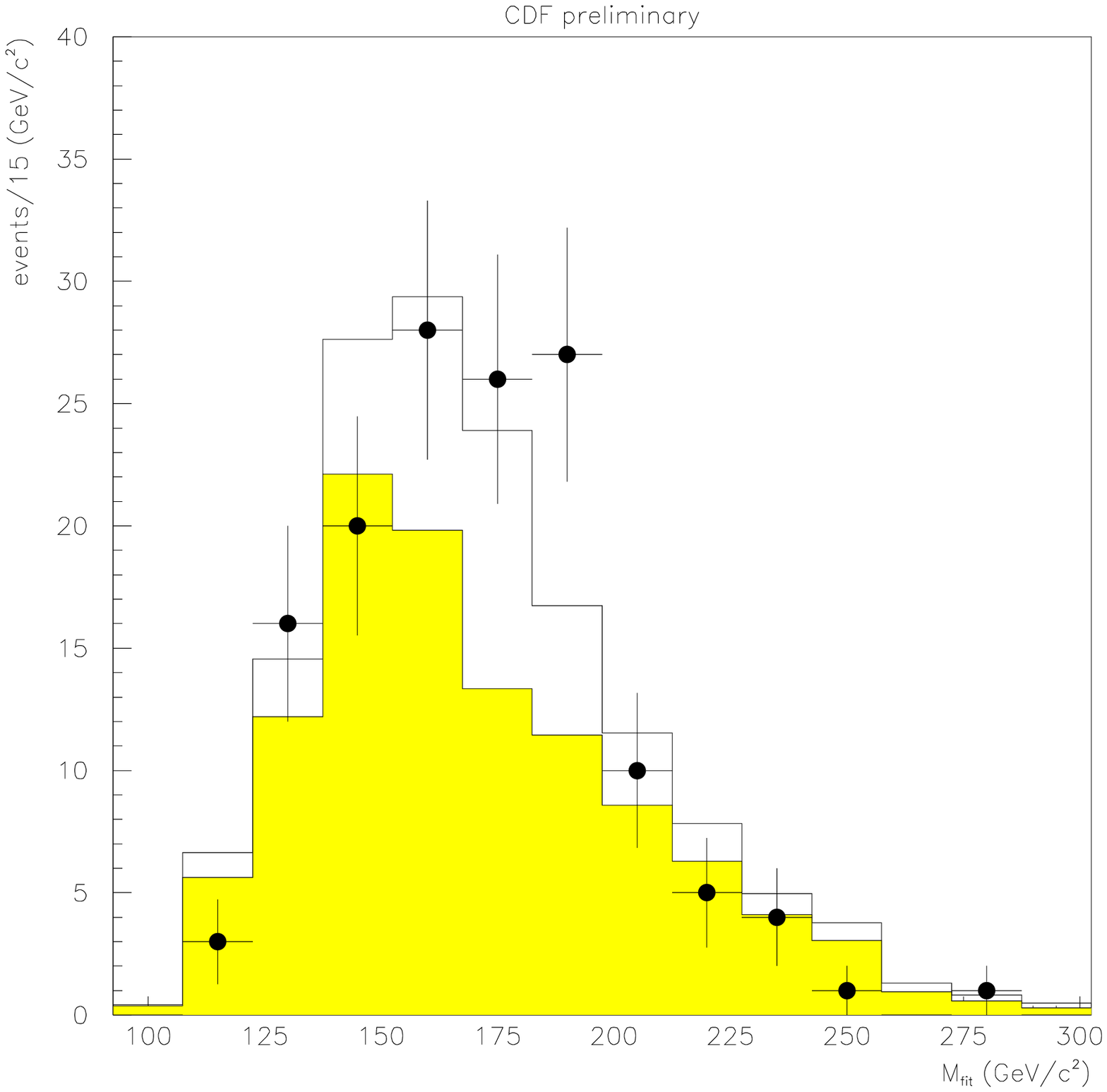}}
\vspace{1.3cm}
\caption{Left (L): Top production cross section as a function of top mass.
One of the theoretical calculations is shown for comparison. 
Right (R): Reconstructed top mass of the $b$--tagged events.}
\label{all_had2}
\end{figure}

\subsection{Cross section evaluation}

The cross section is evaluated, under the assumption of 
$t\bar{t}$ production in the data, using the observed number of candidate 
$b$--tagged events, the expected 
background corrected for the presence of top events in the sample and
the kinematic and $b$--tagging acceptances. 
%, the cross section is evaluated from:
%$\sigma_{t\bar t}=\frac{N_{obs}-N_{exp}'}{\epsilon_{kinematic}\cdot
%\epsilon_{b-tag}\cdot\int {\cal L}dt}$. 
Fig.~\ref{all_had2}(L) shows the $t\bar{t}$  
production cross section as a function of the top mass. The solid line 
represents the all hadronic data measurement, the dashed lines the $\pm$ 
global uncertainty. The main contributions to the systematic uncertainty 
come from the jet energy scale and from the choice of the fragmentation and 
gluon radiation modeling. The result is consistent with the CDF 
measurements from leptonic channels.  
Using the acceptance of the kinematic requirements for 
$M_{top}$ = 175 $GeV/c^2$ we measure the production cross section to be
$\sigma_{t\bar{t}}^{HAD}$ = 10.7$^{+7.6}_{-4.0}pb$. 

\subsection{Mass determination}

A kinematic reconstruction of the mass is applied to a sub--sample of 
events with 6 or more jets, selected with a looser kinematic set of cuts.
A maximum likelihood method is used to extract 
the top mass value. The minimum is found at $M_{top}$ = 187 $\pm$ 8 (stat.) 
$^{+13}_{-12}$(syst.) $GeV/c^2$. Fig.~\ref{all_had2}(R) shows 
the reconstructed top quark mass for the $b$--tagged events. The black dots 
are the data, the shaded histogram indicates  
the background shape and the white histogram is the sum of background and 
$t\bar t$ contribution, assuming $M_{top}$ = 175 $GeV/c^2$. The background 
is normalized to the estimate from the tag rate. 
For more details on the mass determination we 
refer to another contribution to this Conference~\cite{rolli}.

\section{Double tags study}

In the same channel a second analysis has been performed, with a different 
approach to the isolation of a $t\bar{t}$ signal. Starting from the same 
sample, we require the presence of a second $b$ tag in the events 
satisfying $\sum$ $E_T$ $\ge$ 300 $GeV$. All the physics processes that can 
result in $\ge$ 2 heavy flavor quarks in the final state have been studied. 
The dominant sources of background are mistags and QCD heavy flavor pair 
production. In 
four jet events, where the top contribution is expected to be negligible, 
these two processes describe well the data. In the events with a higher jet 
multiplicity, an excess is observed which requires the presence of a 
$t\bar{t}$ component in order to be explained. To enhance the significance 
of the excess  we use three kinematic variables that have different 
distributions for signal and background: the azimuth difference between the 
two $b$ tags ($\Delta\phi(b\bar{b})$), the separation in $\eta - \phi$ 
space ($\Delta R(b\bar{b})$) and the invariant mass of the two $b$ jets. 
From a combined fit of these distributions we obtain a $t\bar{t}$ 
production cross section $\sigma_{t\bar{t}}^{HAD}=7.1^{+4.8}_{-3.5}~pb$ 
(preliminary 
uncertainty estimate), consistent with the value found in the previous 
analysis. 
   
\section{$\tau$ dilepton channel}

The first motivation for investigating this channel is to test the Standard 
Model prediction. In addition, opening a new channel is a way to increase 
the acceptance for top quark decay, expecially in the dilepton channel, 
which has a small branching ratio. 
 
The hadronic branching ratio of $\tau$ decays is $\approx$ 64\%. Each 
$\tau$ decay involves an undetectable neutrino, which decreases the 
kinematic acceptance for the $\tau$ decay products. The $\tau$ 
identification is not as efficient as for $e$ and $\mu$, if we want to keep 
under control the background from jets. The total acceptance for 
$\tau$'s is therefore smaller than for standard dileptons.

In the past, an algorithm to identify hadronically decaying $\tau$'s was 
used to study lepton universality in $W\rightarrow\tau\nu$ decays
~\cite{roodman}. Top dileptons with hadronic $\tau$'s were previously 
searched for in a smaller data sample~\cite{io}. Now we present the first 
evidence of a top signal in the $\tau$ dilepton channel
obtained using the full data sample~\cite{michgall}.

\subsection{Event Selection}

The primary lepton is selected as in the standard dilepton analysis
~\cite{kruse}. $Z \rightarrow l^+l^-$ events are removed using tracking and 
calorimeter information. We require at least 2 jets with $E_T$ $\ge$ 10 
$GeV$. Additional background rejection is obtained cutting on the total 
transverse energy $H_T$ of the 
events ($H_T$ $\ge$ 180 $GeV$) and on missing $E_T$ significance
($\sigma_{\ebar_T} = \ebar_T /\sqrt{\Sigma E_T}
(\ebar_T /\sqrt{\Sigma E_T + P_T^{\mu})}\ge~ 3~ GeV^{1/2}$ for $e\tau$ 
($\mu\tau$)). 
Due to the softer $P_T$ spectrum for $\tau$ decay products, 
we increase the acceptance requiring $P_T$($\tau$) $\ge$ 15 $GeV$, $|\eta_{
\tau}|$ $\le$ 1.2. We 
impose the $\tau$ candidate tracks to be isolated and reject tracks 
associated to a calorimeter energy deposition 
consistent with coming from an $e$ or $\mu$. For more details on the 
variables used for $\tau$ identification, we refer to~\cite{michgall}.

\begin{figure}[th]
\centerline{
%\hskip 1.0cm
\epsfysize 9.cm
\epsffile[24 149 549 689]{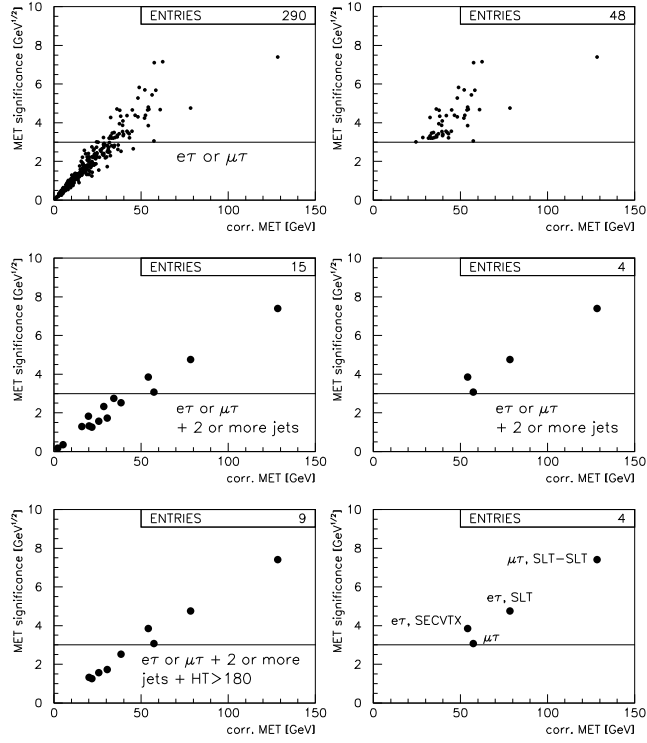}}
\vspace{0.5cm}
\caption{$\sigma_{\ebar_T}$ versus $\ebar_T$ distribution for $\tau$ dilepton 
events. Top: after identifying a primary $e$ ($\mu$) and a $\tau$. Center: 
after requiring $\ge$ 2 jets. Bottom: $\ge$ 2 jets and $H_T$$\ge$ 180 
$GeV$. The line represents the $\sigma_{\ebar_T}$ cut, applied in the plots 
on the right. Events containing 
$b$ tags are labeled $SECVTX$: secondary vertex tag or $SLT$: soft lepton
tag.
(CDF preliminary).}
\label{taus}
\end{figure}

%\vspace{-0.5cm}

\subsection{Backgrounds estimate}

The main background comes from generic jets. There is a small 
probability for a jet to fragment with a low track multiplicity and to be 
identified as a $\tau$ candidate. Leptonically decaying $W$ + $\ge$ 3 jets, 
where one jet is misidentified as a $\tau$, would give a fake $\tau$ 
dilepton event. We obtain the probability for a jet to be wrongly 
identified as a $\tau$ from a sample of generic 
jets, making the assumption that jets in $W$ + $\ge$ 3 jets 
behave the same as generic jets. We parametrize this probability 
as a function of $E_T$(jet).
%the probability for a jet to be wrongly identified as a $\tau$. 
We then apply 
this parametrization to the $W$ + $\ge$ 3 jet sample and
 obtain the background
expectation from fakes. 
The physics background is evaluated using Monte Carlo simulations and data. 

\subsection{Observation in the data}
 
The selection yields 4 candidate events: 2 $e\tau$ and 
2 $\mu\tau$. Three of them are $b$--tagged.
The expected background amounts to 1.96 $\pm$ 0.35 events. 
Fig.~\ref{taus} shows the $\sigma_{\ebar_T}$ as a 
function of the $\ebar_T$ for events passing all the other cuts.
A first measurement of the $t\bar{t}$ production cross section based on 
these events yields $\sigma_{t\bar{t}}^{\tau}$ = 
15.6$^{+18.6}_{-13.2}(stat.)$$pb$. Studies of systematic errors are 
in progress.

\section{Conclusion}

We reported evidence for $t\bar{t}$ production in the all hadronic channel 
and in the $\tau$ dilepton channel. 
%The results are based on 
%110 $pb^{-1}$ of data collected with the CDF detector. 
%Despite the huge background, we observed a signal in the hadronic mode. 
The cross section 
and mass measured in the hadronic mode agree with the values previously 
measured from leptonic 
channels~\cite{kruse}. The observation established in the $\tau$ 
dilepton channel confirms the Standard Model predictions and increases the 
sample of dilepton candidates by including four additional events.

\section*{Acknowledgments}
I would like to thank the organizers of the Conference for giving me the 
opportunity to take part. I am grateful to P. Azzi for her precious help 
in preparing this talk.

\end{document}